\documentclass[a4paper,12pt]{article}

\usepackage{graphicx}
\usepackage{epstopdf}
\usepackage{upgreek}
\usepackage{amsmath}
\usepackage{amssymb}
\usepackage{enumerate}
\usepackage{fullpage}
\usepackage{verbatim}
\usepackage{hyperref}

\pdfoutput = 1

\hypersetup{
    bookmarks=true,         
    unicode=false,          
    pdftoolbar=true,        
    pdfmenubar=true,        
    pdffitwindow=false,     
    pdfstartview={FitH},    
    pdftitle={Supplementary Information for Quantum superposition of a single microwave photon in two different ''colour'' states},    
    pdfauthor={},     
    pdfsubject={Quantum superposition of a single microwave photon in two different ''colour'' states},   
    pdfcreator={},   
    colorlinks=true,       
    linkcolor=red,          
    citecolor=red,        
    filecolor=red,      
    urlcolor=red,          
    pdfpagelayout=TwoPageLeft
}

\renewcommand{\mu}{\upmu}

\title{Quantum superposition of a single microwave photon in two different ``colour'' states}

\author
{ Eva Zakka-Bajjani,$^{1}$ Fran\c{c}ois Nguyen,$^{1}$ Minhyea Lee,$^{2}$ Leila R. Vale,$^{1}$ \\ Raymond W. Simmonds,$^{1}$ Jos\'{e} Aumentado$^{1}$}

\date{}

\begin{document}

\maketitle

\begin{enumerate}
 \item[$^1$] {\it National Institute of Standards and Technology, 325 Broadway, Boulder CO 80305, USA}
 \item[$^2$] {\it Department of Physics, University of Colorado, Boulder, CO 80309, USA}
\end{enumerate}

\begin{abstract}
The ability to coherently couple arbitrary harmonic oscillators in a fully-controlled way is an important tool to process quantum information \cite{haroche2006exploring}. Coupling between quantum harmonic oscillators has previously been demonstrated in several physical systems by use of a two-level system as a mediating element \cite{rauschenbeutel2001controlled,wang2010deterministic}. Direct interaction at the quantum level has only recently been realized by use of resonant coupling between trapped ions \cite{brown2011coupled,harlander2011trapped}. Here we implement a tunable direct coupling between the microwave harmonics of a superconducting resonator by use of parametric frequency conversion \cite{louisell1961quantum,tucker1969quantum}. We accomplish this by coupling the mode currents of two harmonics through a superconducting quantum interference device (SQUID) and modulating its flux at the difference ($\sim$ 7 GHz) of the harmonic frequencies. We deterministically prepare a single-photon Fock state \cite{wallraff2004strong} and coherently manipulate it between multiple modes, effectively controlling it in a superposition of two different ``colours''. This parametric interaction can be described as a beam-splitter-like operation that couples different frequency modes. As such, it could be used to implement linear optical quantum computing protocols \cite{knill2001scheme,milburn2009photons} on-chip \cite{matthews2009manipulation}.
\end{abstract}

The ability to create and manipulate quantum number states in a linear resonator is an important task in cavity quantum electrodynamics (QED) \cite{haroche2006exploring}. Early theory \cite{louisell1961quantum,tucker1969quantum} predicted that parametric frequency conversion could be a way to implement a tunable direct coupling between quantized modes of different energies. Classically, two harmonic oscillators coupled through a time-varying element, modulated at the difference of the resonator frequencies, will periodically exchange energy. At the quantum level, this can be used to swap the quantum states of two harmonic modes. In optics, the efficiency and quantum coherence of frequency up-conversion were demonstrated by use of a pumped nonlinear crystal to couple light at different wavelengths \cite{huang1992observation,tanzilli2005photonic,rakher2010quantum}. However, in these experiments it is challenging to access the state dynamics because strong coupling rates are difficult to obtain \cite{vandevender2004high}. In hybrid mechanical systems, strong parametric coupling, based on frequency conversion, has been recently achieved \cite{gröblacher2009observation,teufel2011circuit}, creating the possibility for the manipulation of quantum states of mesoscopic mechanical resonators \cite{wallquist2010single}.
In superconducting circuits, parametric processes have been used mainly to couple superconducting quantum bits (qubits) at their optimal points \cite{niskanen2007quantum}, or to make quantum-limited microwave amplifiers \cite{yurke1989observation,bergeal465phase,castellanos2008amplification,yamamoto2009flux}, yet little has been done with frequency conversion. Several circuit designs that enable frequency conversion between linear resonators have been proposed \cite{PhysRevLett.104.230502,tian2008parametric,bergeal465phase}. This particular interaction can be combined with the powerful tools already available in circuit QED \cite{wallraff2004strong}, where the creation and the detection of complex quantum states of the field have been demonstrated \cite{hofheinz2009synthesizing}. Here we measure the coherent dynamics of the parametric frequency conversion of a single photon between the first three internal resonant modes of a superconducting cavity, whose state is prepared and read out with a superconducting qubit.

Our circuit consists of a quarter-wave ($\lambda/4$) coplanar waveguide (CPW) resonator terminated to ground via a SQUID (Fig. 1a). At the other end, it is coupled to a 50 $\Omega$ transmission line through a capacitance $C_\textrm{c} $ in order to weakly probe the resonator with microwave reflectometry. At this node we also couple it to a flux-biased phase qubit through an effective capacitance $C_\textrm{g}$. This provides a single photon source and detector for cavity photons.
The SQUID is a superconducting loop asymmetrically intersected by three Josephson junctions.
A ``pump'' line allows both dc ($\Phi_{\textrm{sq}}^{\textrm{dc}}$) and microwave modulation ($\Phi_{ \textrm{sq}}^{\textrm{$\mu$w}}(t)$) of the global flux ($\Phi_{\textrm{sq}}$) in the SQUID loop. In the linear current  regime (Supplementary Information), and for frequencies lower than the plasma frequency, we can model the SQUID as a flux-dependent lumped-element inductor, $L_{\textrm{sq}}(\Phi_{\textrm{sq}})$, modifying one of the boundary conditions of the cavity. Each cavity mode n (assumed lossless in this model) behaves as a flux-tunable harmonic oscillator (Fig. 1b) with a resonance frequency $\nu_{\textrm{n}}(\Phi_{\textrm{sq}}^{\textrm{dc}}) = 1/ 2\pi \sqrt{C_{\textrm{n}}L'_{\textrm{n}} (\Phi_{\textrm{sq}}^{\textrm{dc}}) }$,  where $L'_{\textrm{n}}(\Phi_{\textrm{sq}}^{\textrm{dc}}) = L_{\textrm{n}}+L_{\textrm{sq}}(\Phi_{\textrm{sq}}^{\textrm{dc}})$ and $L_{\textrm{n}}C_{\textrm{n}}$ is the lumped series oscillator describing the modes of the bare cavity  \cite{palacios2008tunable,sandberg2008tuning,yamamoto2009flux}.
Figure 1c shows the spectroscopic data measured as a function of both probe frequency and dc flux $\Phi_{\textrm{sq}}^{\textrm{dc}}$, for the first two modes 0 and 1.
The measurements shown in Fig. 2 and 3 are performed at a flux bias point $\Phi_{\textrm{sq}}^{\textrm{dc}} = -0.37$ $\Phi_0$ (point ``A'' in Fig. 1c), where $\nu_0 = 3.7 $ GHz and $\nu_1 = 10.74 $ GHz (to simplify the notation, the dependence on DC flux will not be specified for the cavity resonances). The $n = 2$ mode cannot be directly probed in our setup, but is expected to be around $ 18 $ GHz. The extracted loaded quality factors are about 9000 for both modes 0 and 1.
The SQUID inductance can be modulated by applying a small microwave flux $\Phi_{ \textrm{sq}}^{\textrm{$\mu$w}}(t) = \delta\Phi_{ \textrm{sq}}^{\textrm{$\mu$w}} \cos (2 \pi \nu_{\textrm{p}} t + \varphi_{\textrm{p}})$, with $\delta\Phi_{ \textrm{sq}}^{\textrm{$\mu$w}} \ll \Phi_0$ at a pump frequency $\nu_{\textrm{p}}$. To first order in flux, the SQUID inductance is (the phase of the pump is defined modulo $\pi$, depending on the sign of the first derivative of the SQUID inductance with flux)
\begin{eqnarray}\label{inductance}
    &L_{\textrm{sq}}(\Phi_{\textrm{sq}}^{\textrm{dc}} + \Phi_{\textrm{sq}}^{\mu\textrm{w}}(t)) = L_{\textrm{sq}}(\Phi_{\textrm{sq}}^{\textrm{dc}})+ \textrm{ } \left| \delta L\right|\textrm{ }  \cos (2 \pi \nu_{\textrm{p}} \textrm{ } t + \varphi_\textrm{p}), & \\
   \textrm{with } &\delta L = \delta L(\Phi_{\textrm{sq}}^{\textrm{dc}}, \delta\Phi_{ \textrm{sq}}^{\textrm{$\mu$w}}) = \frac{\partial L_{\textrm{sq}}}{\partial \Phi}  \bigg|_{\Phi_{\textrm{sq}}^{\textrm{dc}}} \delta\Phi_{ \textrm{sq}}^{\textrm{$\mu$w}}.& \nonumber
\end{eqnarray}
When $\nu_{\textrm{p}} = \nu_{\textrm{m}}-\nu_{\textrm{n}}$ (by convention $m>n$), frequency conversion is induced between modes n and m \cite{louisell1960coupled,tucker1969quantum,louisell1961quantum}. Since, in a usual $\lambda/4$ cavity, the mode frequencies are $\nu_{n}^{\lambda/4} = (2n+1)\nu_{0}^{\lambda/4}$, a pump at $2\nu_{0}^{\lambda/4}$ can also couple other neighboring modes by conversion, as well as induce degenerate parametric amplification of mode 0 \cite{louisell1960coupled}. In order to couple only two selected modes by frequency conversion (for a given pump frequency), we modify the cavity dispersion relation by slightly varying the characteristic impedance along the CPW resonator. This ensures that the frequencies $\nu_{\textrm{n}}$ are not equally spaced, satisfying the conditions: $|\nu_{2}-\nu_{1}|-|\nu_{1}-\nu_{0}|, |\nu_{1}-\nu_{0}| - 2\nu_{0} >> \Delta \nu_{0}, \Delta \nu_{1}, \Delta \nu_{2}$, where $\Delta \nu_{n}$ is the bandwidth of mode n (we measure $\Delta\nu_{0,1} \lesssim 2$ MHz). This allows us to restrict our description to a two-mode manifold. Following the usual treatment for two parametrically coupled oscillators \cite{louisell1960coupled}, one can write the classical equations of motion of the two normalized normal mode amplitudes $a_{\textrm{k}} = \sqrt{L'_{\textrm{k}}/h \nu_\textrm{k}} \textrm{ }(\dot{I}_{\textrm{k}}/( 2 i \pi  \nu_{\textrm{k}}) + I_{\textrm{k}})$, where $I_{\textrm{k}}$  is the current in mode k (k$\in \{\textrm{n, m}\}$), coupled by a quasi-resonant pump of frequency $\nu_{\textrm{p}}=\nu_{\textrm{m}}-\nu_{\textrm{n}}+\Delta \nu_{\textrm{p}}$ ($\Delta \nu_{\textrm{p}}\ll\nu_{\textrm{m}},\nu_{\textrm{n}}$). Keeping the resonant terms for each mode and invoking the rotating wave approximation, one has
\begin{eqnarray}\label{motionequation}
\dot{a}_{\textrm{n}}(t)/2 \pi & = & -i \textrm{ } \nu_{\textrm{n}} a_{\textrm{n}}(t) -i \textrm{ }  g_{\textrm{mn}} e^{i (2 \pi  \nu_\textrm{p} t + \varphi_\textrm{p})} a_{\textrm{m}}(t) \\
\dot{a}_{\textrm{m}}(t)/2 \pi & = & -i \textrm{ } \nu_{\textrm{m}} a_{\textrm{m}}(t) -i \textrm{ } g_{\textrm{mn}} e^{-i (2 \pi  \nu_\textrm{p} t + \varphi_\textrm{p})} a_{\textrm{n}}(t),\nonumber
\end{eqnarray}
where $g_{\textrm{mn}} = \frac{1}{2} \textrm{ } \textrm{ } \sqrt{\nu_{\textrm{m}}\nu_{\textrm{n}}} \left|\delta L\right|/\sqrt{L'_{\textrm{m}}L'_{\textrm{n}}}$ is the parametric coupling frequency. For simplicity, we neglect its dependence on the pump detuning. The quantum model is derived by identifying Eq. 2 with Heisenberg's equations of motion for the annihilation operators $\hat{a}_{\textrm{n,m}}(t)$ (the pump field is assumed to be strong enough to be described classically). The state associated with each harmonic mode of the cavity can be described in terms of quantized microwave excitations, {\it i.e.} photons. In the doubly rotating frame defined by the unitary transformation $e^{2 i \pi [(\nu_\textrm{n} - \Delta \nu_\textrm{p}/2) \hat{a}^\dagger_\textrm{n} \hat{a}_\textrm{n} + (\nu_\textrm{m} + \Delta \nu_\textrm{p}/2) \hat{a}^\dagger_\textrm{m} \hat{a}_\textrm{m}] t}$, the Hamiltonian, in the Schr$\ddot{\textrm{o}}$dinger representation, is
\begin{eqnarray}\label{hamiltonian}
\hat{H}_I = h \frac{\Delta \nu_\textrm{p}}{2} (\hat{a}^\dagger_{\textrm{n}} \hat{a}_{\textrm{n}} - \hat{a}^\dagger_{\textrm{m}} \hat{a}_{\textrm{m}}) + h g_\textrm{mn} (\hat{a}^\dagger_{\textrm{n}} \hat{a}_{\textrm{m}} e^{i \varphi_\textrm{\scriptsize p}} + \hat{a}_{\textrm{n}} \hat{a}^\dagger_{\textrm{m}} e^{-i \varphi_\textrm{p}}).
\end{eqnarray}
When $\Delta \nu_{\textrm{p}} = 0$, $\hat{H}_I$ describes a generalized beam-splitter operation between modes of different frequencies, which preserves the total number of photons \cite{haroche2006exploring}. The effective transparency is modulated by the parametric interaction duration $\Delta t_{\textrm{p}}$. In particular, the complete swap of a given initial state from one mode to the other is achieved for a pump $\pi$-pulse of duration $\Delta t_{\textrm{p}} = 1/(4 g_{\textrm{mn}})$.

Figures 2a and 2b show two separate sets of spectroscopic data probing the parametric coupling between modes n = 0 and m = 1. They are plotted as a function of $\nu_{\textrm{p}}$ and driven at three different pump amplitudes $A_{\textrm{p}, \mu\textrm{w}}$ (measured at the output of the microwave generator). The probe frequency $\nu_{0,1}^\textrm{probe}$ is swept through $\nu_0$ (Fig. 2a), and $\nu_1$ (Fig. 2b). These data show well-resolved normal-mode splittings centered at $\nu_\textrm{p}=\nu_1-\nu_0 \approx 7.043$ GHz, characteristic of the strong-coupling regime. In other words, the parametric coupling rate $g_{10}$ is larger than the bandwidths of the modes. The extracted $g_{10}$ rate is linearly dependent on pump amplitude yielding up to 20 MHz coupling for a flux modulation $\delta\Phi_{ \textrm{sq}}^{\textrm{$\mu$w}}/\Phi_0 \sim 2$ $\%$. As the SQUID inductance is slightly rectified by the pump modulation, we observe a small shift of the resonance frequencies, leading to a shift of the splitting centers (Supplementary Information). The ``quasi-resonant pump'' condition, $\nu_\textrm{p} = \nu_1-\nu_0^\textrm{probe}$ ($\nu_\textrm{p} = \nu_0+\nu_1^\textrm{probe}$), describes the asymptotic behaviour at small pump detuning. We can also couple modes 1 and 2 (Fig. 2c). Because we designed the resonator to shift the harmonic mode frequencies, this requires a pump frequency $\sim 250 $ MHz higher than that needed for the 0$\leftrightarrow$1 conversion. In addition, the 1$\leftrightarrow$2 conversion rate is consistent with expectations from our simple model (Eq. 2), $g_{21} \sim \sqrt{5} g_{10}$. Since $g_{10}, g_{21}$ are small enough compared to $\left| \nu_2-\nu_1\right| - \left| \nu_1-\nu_0\right|$, we can, in fact, address these dual-mode manifolds separately.

As an illustration of the circuit efficiency at the quantum level, we realize the conversion oscillations of a one-photon Fock state between modes 0 and 1 (1 and 2).
We prepare and measure the state of mode 1 using a superconducting phase qubit following the method outlined in \cite{sillanpää2007coherent,hofheinz2009synthesizing}. The qubit is well described by a two-level system, with a ground state $\left| \textrm{g} \right>$ and an excited state $\left| \textrm{e} \right>$. Their energy difference defines the qubit frequency $\nu_\textrm{q}$, which we can tune from 7 to 12 GHz by use of an external flux $\Phi_\textrm{q}$ (Supplementary Information). The qubit state is manipulated using microwave pulses and  readout with a dc SQUID.
By applying a fast flux pulse to the qubit, the $\left| \textrm{e} \right>$ state preferentially tunnels to a different flux state, which is indicated by a shift in the dc SQUID critical current.
We measure a relaxation time $T_1^\textrm{q} \approx 100$ ns, and a qubit-resonator coupling rate $g_{\textrm{q1}} = $ 32 MHz.
We load a single-photon Fock state in mode 1 by preparing the qubit in state $\left| e \right>$ at $\nu_\textrm{q} = 10.116$ GHz, bringing it in resonance with the cavity for a duration $T_\pi = 1/4 g_{\textrm{q1}} = 8$ ns, and shifting it back out of resonance. This prepares the cavity state $\left| 0_0 1_1 \right>$ (the subscript denotes the mode number).
After applying a pump pulse of duration $\Delta t_\textrm{p}$, the state of mode 1 is transferred back to the qubit by bringing it again in resonance with the cavity during a duration $T_\pi$. Finally we measure the state of the qubit.
Figure 2d (2e) shows single photon conversion oscillations between modes 1 and 0 (resp. 1 and 2) as a function of $\Delta t_\textrm{p}$ and $\nu_{\textrm{p}}$.
At zero pump detuning, the frequency of the oscillations is the lowest, equal to the splitting strength measured in spectroscopies 2(a b c), and their amplitude is at their maxima.
The Fourier transforms of these data are computed in Fig. 2f (Fig. 2g) and exhibit the expected hyperbolic dependence $\sqrt{(2g_{10})^2+{\Delta\nu_{p}}^2}$ $\left(\sqrt{(2g_{21})^2+{\Delta\nu_{p}}^2}\right)$, when the initial state is a single-photon Fock state in one mode.
The amplitude of these oscillations follows an exponential decay, with a characteristic time $T_{10}^{\textrm{Rabi}}$. Its measured value is $T_{10}^{\textrm{Rabi}} \approx 180$ ns ($T_{21}^{\textrm{Rabi}} \approx 100$ ns) compatible with the harmonic mean of the measured relaxation times $T_{0}^{\textrm{Decay}} \approx 370$ ns and $T_{1}^{\textrm{Decay}} \approx 140$ ns of both modes.

We measure the coherence of this parametric process by performing a Ramsey interference experiment with a photon ``shared'' between modes 0 and 1, as a function of the pump detuning, with a pump amplitude corresponding to $g_{10} = 20 $ MHz. After preparing $\left| 0_0 1_1 \right>$, we apply two phase-coherent pump pulses of $6$ ns duration (corresponding to a $\pi/2$ pulse at $\Delta \nu_\textrm{p} = 0$), separated by a variable delay $\Delta t_\textrm{r}$, and then measure the final state of mode 1. The first pulse prepares a superposition of states $\left| 0_0 1_1 \right>$ and $\left| 1_0 0_1 \right>$. During the free evolution, these two states acquire opposite phases differing by $2\pi\Delta \nu_\textrm{p} \Delta t_\textrm{r}$. The second pulse combines these states, resulting in interference oscillations with a period equal to the inverse of the detuning. We verify this experimentally (Fig. 3). The amplitude of the oscillations decays on a characteristic timescale $T_{10}^{\textrm{Ramsey}} \approx 190$ ns, which indicates that the decoherence is mainly due to relaxation.

To further confirm the circuit model, we measure the $1\leftrightarrow0$ conversion rate (Fig. 4a) as a function of the pump amplitude, for different $\Phi_{\textrm{sq}}^\textrm{dc}$ (points ``A, B, C, D, E'' on Fig. 1c). For the flux bias range that we explore ($[-0.4,-0.3]\Phi_0$), the coupling rates $g_{10}$ are linearly dependent on the experimentally accessible pump amplitudes.
The variation of the conversion efficiency $\partial g_{10}/\partial A_{\textrm{p},\mu\textrm{w}}$ with the flux is also in good agreement with our SQUID inductor model and its simple expansion in Eq. 1, and is consistent with measured values of the pump line attenuation and dc mutual inductance to pump line (Fig. 4b).

To conclude, we have demonstrated that parametric frequency conversion is an effective way to coherently manipulate a single microwave-photon Fock state in frequency space. The dynamics are accurately described by a generalized beam-splitter interaction familiar from quantum optics. Combined with a phase shift operation, which can be implemented by fast shifts in SQUID bias flux, this system could be used as a novel linear optical quantum bit based on microwave resonator modes. Furthermore, straightforward technical improvements will enable the manipulation of multi-photon states. Lastly, this device also offers the opportunity to explore other parametric interactions such as amplification with similarly strong interaction rates.

\bibliography{scibib}
\bibliographystyle{naturemag}

\begin{enumerate}
\item[] {\bf Acknowledgments}
We thank N. Bergren, L. Ranzani for technical help, and J. Park, F. Altomare, L. Spietz for valuable input. This paper is a contribution by the National Institute of Standards and Technology and not subject to US copyright.
\item[] {\bf Author Contribution}
E.Z-B. and F.N. designed the experiment, built the measurement set-up and performed the measurements. M.L., R.W.S., J.A. contributed to the experimental design. L.R.V. contributed to the fabrication process development. J.A. conceived the experiment and supervised the project. All authors participated to the sample fabrication, to the writing of the manuscript and to the data analysis.
\item[] {\bf Author Information}
Correspondence and requests for materials should be addressed to E.Z-B. (eva.zakka-bajjani@nist.gov), F.N. (francois.nguyen@nist.gov), or J.A. (jose.aumentado@nist.gov).
\end{enumerate}

\clearpage

\renewcommand{\caption}[2]{\textbf{Figure {#1}} $|$ {#2}}

\begin{figure}
\caption{1}{{\bf Device description and spectroscopy.} {\bf a}, The ``Superconducting Parametric Beam Splitter" (SPBS) is a $\lambda/4$ CPW resonator whose boundary condition is varied by a SQUID. The cavity has a total length of 7.57 mm and is composed of two $\lambda/8$ sections with characteristic impedances $Z_{1} = $ 50 $\Omega$ and $Z_{2} = $ 46 $\Omega$. The SQUID is intersected by three junctions of critical currents $I_{\textrm{c},1} \approx 0.21$ $\mu$A, $I_{\textrm{c},2} \approx 0.41$ $\mu$A, and $I_{\textrm{c},3} \approx 0.88$ $\mu$A. The flux through the SQUID is dc-biased and microwave-modulated via an inductively coupled pump line. To allow reflectometry measurements, the SPBS is coupled to a 50 $\Omega$ transmission line through a small capacitance $C_{\textrm{c}}\approx$ 1.8 fF. A phase qubit is capacitively coupled to the SPBS with a strength $g_{\textrm{q}1}=$ 32 MHz (Supplementary Information). We manipulate the qubit state with microwave pulses and tune its frequency with a flux bias line. Measurements are performed in a dilution refrigerator operated at 35 mK, which makes thermal noise negligible at our working frequencies.
{\bf b}, SPBS simple model. Each cavity mode n is described by the equivalent $L_{\textrm{n}}C_{\textrm{n}}$ of the $\lambda/4$ CPW n$^{\textrm{th}}$ harmonic in series with the SQUID tunable inductor. {\bf(c)} Measured reflection coefficient $\left|S_{11}\right|$ on $C_{\textrm{c}}$ as a function of both dc-flux in the SQUID and probe frequency.}
\end{figure}

\begin{figure}[ht]
\caption{2}{{\bf Spectroscopy of parametrically coupled cavity modes and one-photon Rabi-swap oscillations.} {\bf a, b, c}, Spectroscopic data of modes 0 and 1, in presence of a microwave pump drive on the cavity SQUID, at $\Phi_{\textrm{sq}}^{\textrm{dc}} = $ -0.37 $\Phi_{0}$ (point ``A'' on Fig. 1c). $\left|S_{11}\right|$ is plotted in colour scale as a function of both probe and pump frequencies, when modes 0 and 1 ({\bf a}, {\bf b}), or 1 and 2 ({\bf c}), are distinctly coupled. Parametric interaction enables a strong coupling between harmonics that are 7 GHz detuned, resulting in an avoided crossing between the two eigenmodes of the system (normal-mode splitting). The dashed black lines outline the asymptotic behaviour. The coupling rate $g_{10}$ depends linearly on the pump amplitude $A_{\textrm{p}, \mu\textrm{w}}$. For the same pump amplitude, the coupling rate $g_{21}$ between modes 1 and 2 is about $\sqrt{5}g_{10}$. We notice the presence of an unexplained parametrically coupled resonance at $\nu_\textrm{p} = 7.35$ GHz. {\bf d, e}, Single-photon conversion oscillations. The colour scale encodes the tunneling probability $P_{\textrm{t}}$ of the qubit state as a function of both pump pulse duration and frequency. When the qubit is in the ground state, $P_{\textrm{t}}$ is set to $\approx25$ $\%$. This probability increases linearly with the excited state occupancy. When the qubit is prepared in the excited state, $P_{\textrm{t}}$ is $\approx50$ $\%$. {\bf f, g}, Fourier transform of the time-domain measurements showing the hyperbolic dependence between the coupling rate and the pump detuning. The theoretical curve is superposed in dotted line.}
\end{figure}

\begin{figure}[ht]
\caption{3}{{\bf Single-photon Ramsey interferences.} {\bf a}, Ramsey-fringes experiment with a single photon split between modes 0 and 1, at $\Phi_{\textrm{sq}}^{\textrm{dc}} = $ -0.37 $\Phi_{0}$ and $A_{\textrm{p}, \mu\textrm{w}} = $ 0.39 V at the microwave generator output. The cavity is initially prepared in the state $\left|0_{0}1_{1}\right>$. Two phase-coherent ``$\pi /2$'' pump pulses, separated by a variable delay $\Delta t_{\textrm{r}}$ are then applied. The state of mode 1 is finally measured with the qubit. The tunneling probability $P_{\textrm{t}}$ is plotted as a function of both the pump frequency and the delay $\Delta t_{\textrm{r}}$. {\bf b}, Fourier transform of the time-domain data. When starting with a single-photon Fock state, the frequency of the Ramsey oscillations is equal to the pump detuning $\Delta \nu_{\textrm{p}}$.}
\end{figure}

\begin{figure}[ht]
\caption{4}{{\bf Dependence of the coupling rate with the SQUID flux.} {\bf a} Parametric coupling frequency $g_{10}$ between modes 0 and 1 as a function of the pump amplitude $A_{\textrm{p}, \mu\textrm{w}}$, for the various operating fluxes A, B, C, D, E shown on Fig. 1c. In the explored flux and pump amplitude ranges, $g_{10}$ depends linearly on $A_{\textrm{p}, \mu\textrm{w}}$. {\bf b} Consistency check of the simple flux expansion of SQUID inductance used in our model (Eq. 1). For a given flux, we extract the conversion efficiency $\partial g_{10}/\partial A_{\textrm{p}, \mu\textrm{w}}$ from the measurements in Fig. 4a. This is compared to a theory curve of $\alpha\left|\partial L_{J} / \partial \Phi \right|$, where $\left| \partial L_{J} / \partial \Phi \right|$ is the numerically calculated first derivative with flux of the SQUID inductance , and with $\alpha = \frac{ \sqrt{\nu_{0}\nu_{1}} }{ 2 \sqrt{L_{0}'L_{1}'}} M_{\textrm{sq}} \textrm{ } \kappa$, where $M_{\textrm{sq}}$ is the mutual coupling inductance between the pump line and the SQUID, and $\kappa$ is the attenuation of the pump line. We set the value of the $\frac{\sqrt{\nu_{0}\nu_{1}}}{2\sqrt{L_{0}'L_{1}'}}$ prefactor to the one at point a, and we assume that $M_{\textrm{sq}}$ is equal to the dc value (1.2 pH). $\kappa$ is the only adjustable parameter and is in good agreement, within 2 dB, with the attenuation measured at room temperature.}
\end{figure}

\clearpage

\thispagestyle{empty}
\begin{figure}
\begin{center}
\includegraphics[scale = 0.5]{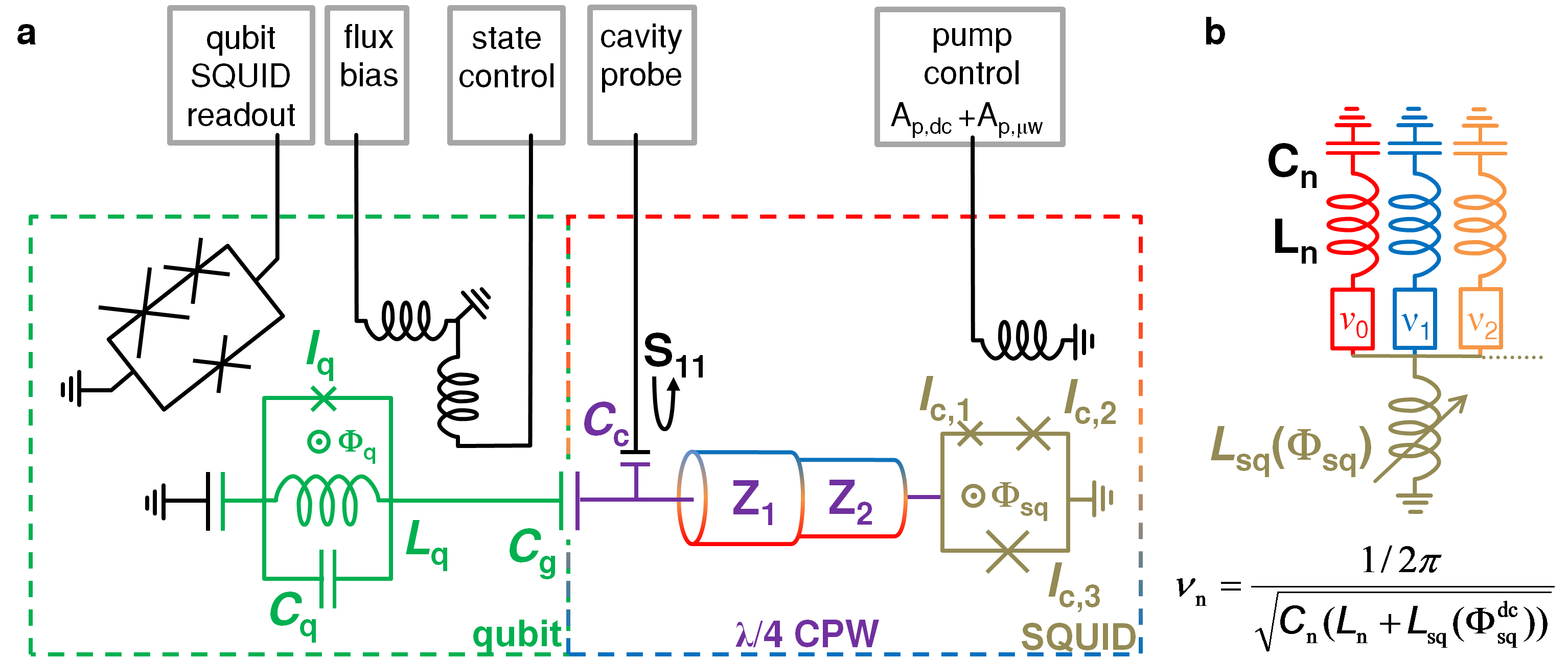}
\end{center}
\end{figure}
\begin{figure}[ht]
\begin{center}
\includegraphics[scale = 3]{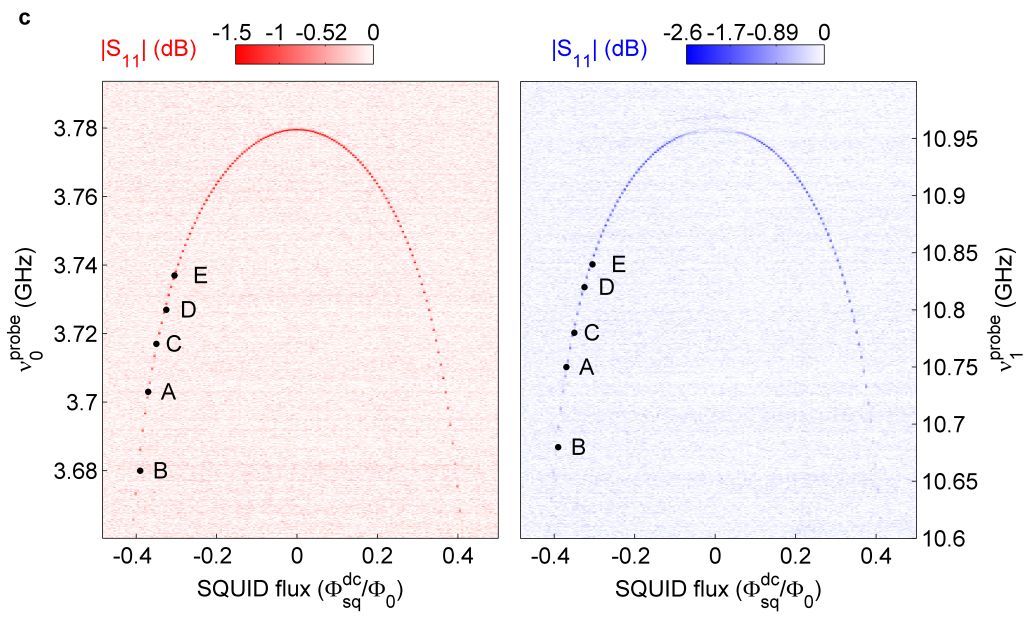}
\end{center}
\end{figure}

\clearpage
\thispagestyle{empty}
\begin{figure}[ht]
\begin{center}
\includegraphics[scale = 0.7]{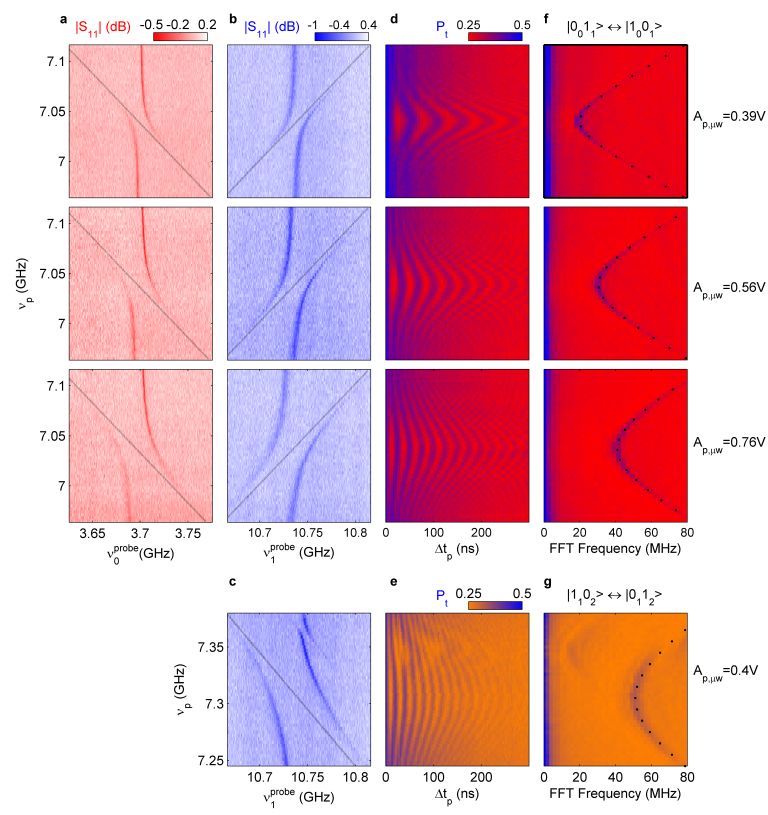}
\end{center}
\end{figure}

\clearpage
\thispagestyle{empty}
\begin{figure}[ht]
\begin{center}
\includegraphics[scale = 3]{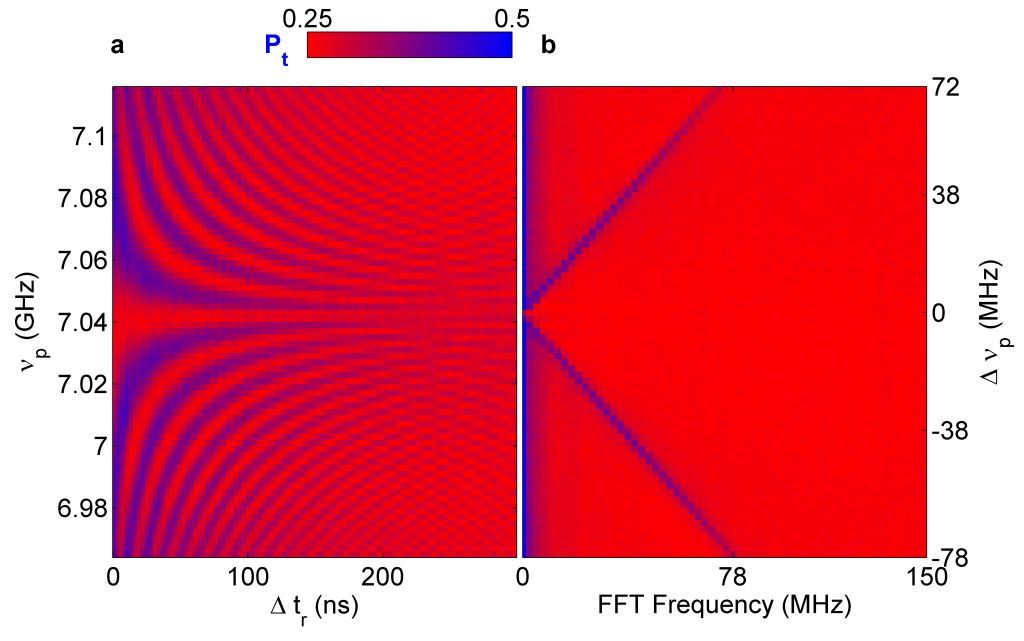}
\end{center}
\end{figure}

\clearpage
\thispagestyle{empty}
\begin{figure}[ht]
\begin{center}
\includegraphics[scale = 0.5]{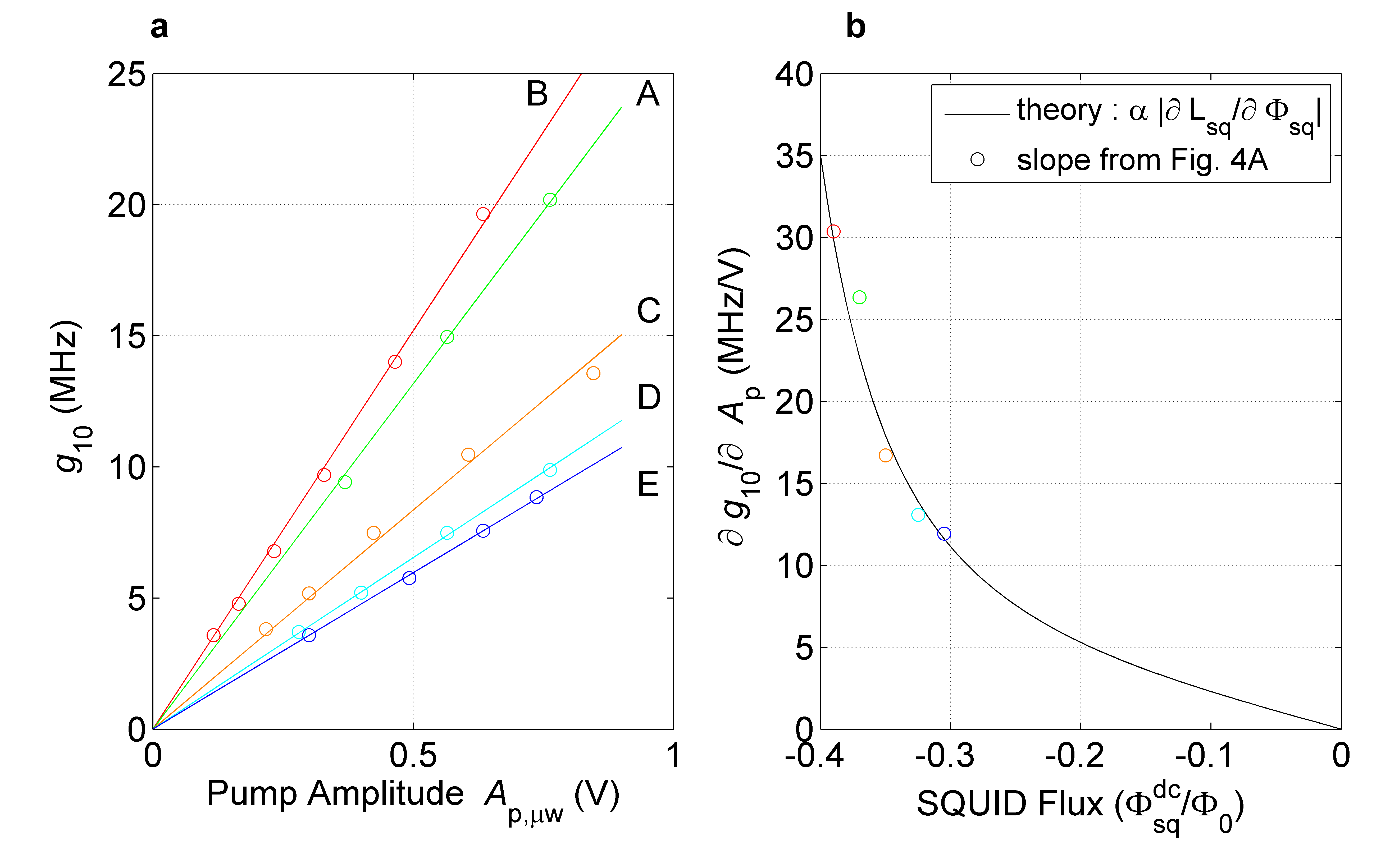}
\end{center}
\end{figure}

\clearpage
\renewcommand{\caption}[2]{\textbf{Supplementary Figure {#1}} $|$ {#2}}
\renewcommand{\thepage}{}

\begin{center}
{\bf {\Large Supplementary Information for `Quantum superposition of a single microwave photon in two different ``colour'' states'}}
\end{center}

\subsubsection*{Methods - Sample fabrication}
The device is fabricated on a 300 $\mu$m thick sapphire substrate by use of standard optical lithographic techniques and liftoff of e-gun deposited metal. First, a 150 nm aluminum base electrode layer is fabricated, providing the interconnects of the subsequent metal layer elements (cavity, qubit, SQUID). To produce the dielectric crossovers for the gradiometric coils, we deposit a 200 nm thick SiO2 layer by PECVD and pattern it by reactive ion etching. We then pattern and deposit a second layer of aluminum which defines the qubit, the readout dc-SQUID and the SPBS (cavity and SQUID wires). Finally we pattern the Josephson junctions of the phase qubit, readout dc-SQUID, and cavity SQUID. After ion-milling the previous aluminum layer for good electrical contact, we do a standard double-angle deposition of aluminum with an intermediate oxidation step. The resulting critical current density is about 1.1 $\mu$A/$\mu\textrm{m}^2$. The chip is wire-bonded to a microwave circuit board that is integrated in a copper sample box. It is anchored to the mixing chamber of a dilution refrigerator operated at a temperature of 35 mK.

\subsubsection*{Methods - Measurement setup}
 The experimental setup is shown on Supplementary Figure 1. To make reflectometry measurements, the cavity is connected to a 50 $\Omega$ transmission line through a small capacitance $C_{\textrm{c}}$ = 1.8 fF. Incoming and outgoing signals are separated with a 20 dB directional coupler. The total attenuation of the input line is about -103 dB. The outgoing signal is amplified by two HEMT amplifiers (0.4-11 GHz, +37 dB gain) at 4 K, followed by a room-temperature amplifier (+20 dB gain). Two wideband isolators (4-12 GHz bandwidth) are placed between the output of the directional coupler and the HEMT amplifiers (one at base temperature, and one at 4 K) to protect the device from nonequilibrium noise. The $S_{11}$ measurement is performed with a network analyzer. This setup allows us to measure the resonator at low intracavity power corresponding to an average of one photon. The pump line is connected to a bias tee, which allows us to modulate both the dc and microwave flux of the SQUID in the cavity. The dc input of the bias-tee is in series with a 10 k$\Omega$ resistor at 4 K and a room-temperature voltage source. We have inserted copper-powder filters at 4 K and base temperature to low-pass filter at a few megahertz. The microwave pump signal is routed through the coupled port of a 20 dB directional coupler connected to the microwave port of the bias-tee. This gives us the ability to attenuate the black-body radiation emitted from higher-temperature stages while also dissipating the majority of the pump power at higher-temperature cold stages that can handle the excess heat load (10 dB attenuator at 800 mK and 50~$\Omega$ at 4~K). The qubit loop is inductively coupled to two distinct coils. A heavily filtered line is used to drive slow (a few megahertz) changes to the qubit flux bias while fast control pulses (microwave and fast shift pulses) are driven through a broadband microwave line with 50 dB of attenuation between room temperature and the mixing chamber. These microwave and fast shift pulses are combined at room-temperature by use of a 16 dB directional coupler. As in other phase qubit measurements $[\hyperlink{31}{31}]$, we read out the qubit by measuring the switching current of an inductively coupled dc SQUID. The readout SQUID is current-biased by use of a 1 k$\Omega$ resistor in series with an arbitrary waveform generator (AWG). The voltage that develops across the SQUID when it switches propagates along the readout line through a 10 k$\Omega$ resistor at 4 K, is amplified by a room-temperature high-impedance low-noise amplifier and finally digitized. The switching probability is computed by averaging the value of the switching current over 5000 events. We used a 5 kHz repetition rate ensuring that the qubit has relaxed to its ground state and that quasiparticles generated by the readout SQUID switching have equilibrated between each measurement. The microwave pulses for the qubit and pump control are generated by mixing microwave signals with ``dc'' pulse sequences generated by an AWG using vector synthesizers. The qubit is prepared in its excited  state with a trapezoidal shaped pulse of 18 ns duration (rise/fall times are 5 ns). The measurement visibility, \emph{i.e.}, the difference between the tunneling probabilities for the $\left|g\right>$ and $\left|e\right>$ states is $\sim 25$ $\%$.
\begin{figure}[ht]
\begin{center}
\includegraphics[scale = 0.5]{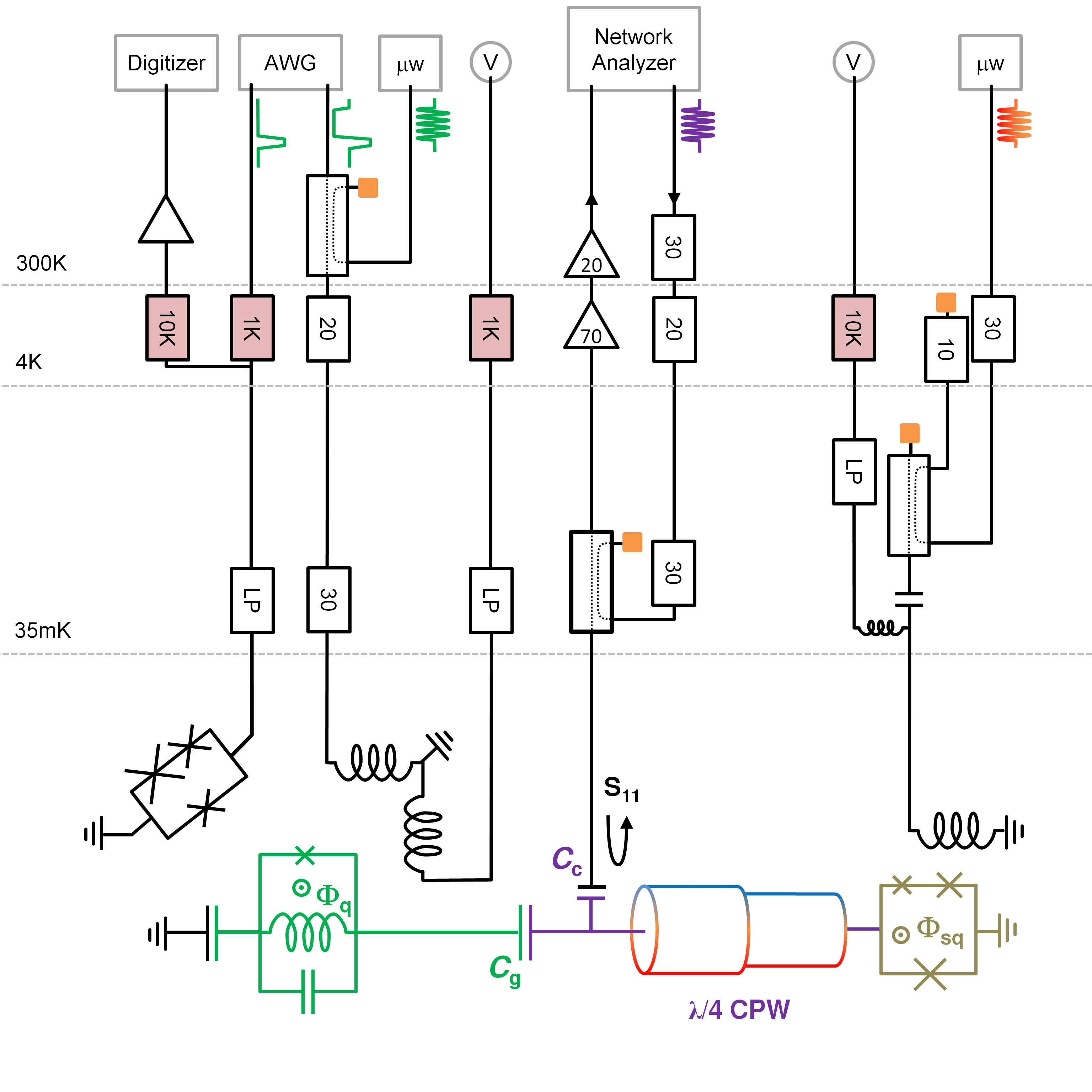}
\end{center}
\caption{1}{{\bf Experimental setup.} Full-coloured blocks represent bias resistors, white ones attenuators, orange ones 50 $\Omega$ loads. LP boxes denote the copper powder filters.}
\end{figure}

\subsubsection*{Methods - Phase qubit characteristics, preparation, and readout}
The phase qubit used in this work has the following design circuit parameters : $L_{\textrm{q}} \approx $ 1.2 nH, $C_{\textrm{q}} \approx $ 0.32 pF, $I_{\textrm{q}}\approx $ 0.4 $\mu$A. In comparison with typical phase qubits $[\hyperlink{32}{32}]$, the hysteresis parameter $\beta_{\textrm{L}}=L_{\textrm{q}}I_{\textrm{q}}/\varphi_{0} \approx $ 1.4, ($\varphi_{0}=\hbar/2e$ is the reduced magnetic quantum flux), is relatively small. For the experiment, the qubit is prepared at $\nu_{\textrm{q}}^{\textrm{off}} = $ 10.116 GHz (noted as ``Off'' on Supplementary Figure 2), which ensures being far detuned from the cavity frequency. At this point, the qubit potential is a single well. The transition frequency between the first and the second excited state is 100 MHz lower than $\nu_{\textrm{q}}^{\textrm{off}}$. We measure a relaxation time $T_{1}^{\textrm{q}} \approx $ 100 ns and a decoherence time $T_{2}^{\textrm{q}} \approx $ 60 ns.

\begin{figure}[ht]
\begin{center}
\includegraphics[scale = 0.5]{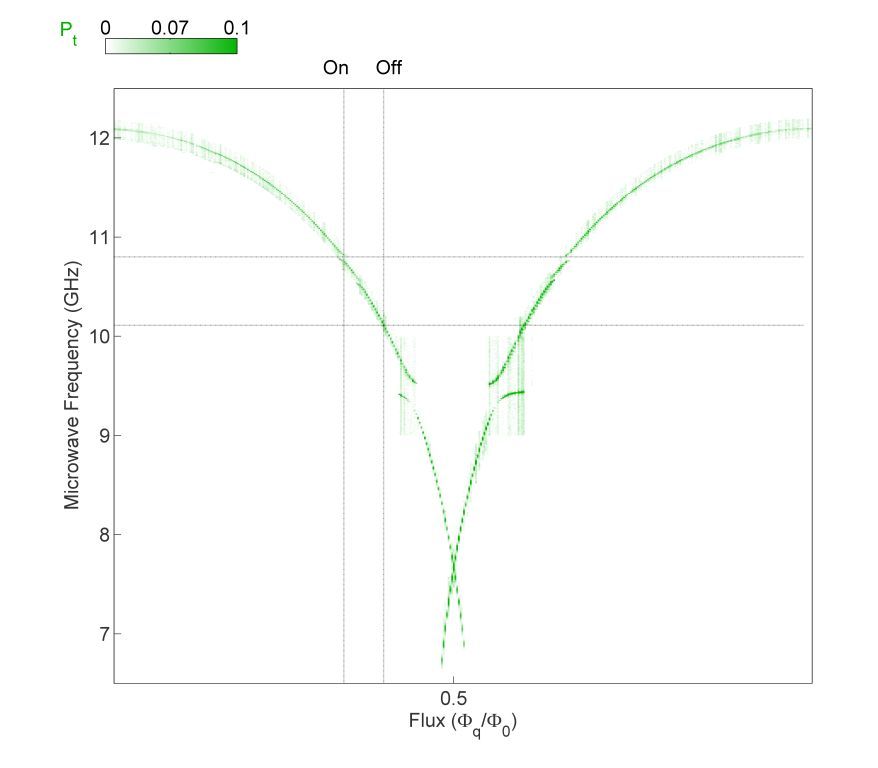}
\end{center}
\caption{2}{{\bf Qubit spectroscopy.} The colour scale encodes the tunneling probability of the qubit state as a function of the probe microwave frequency and the flux through the qubit. The $\approx 9.5 $ GHz and the $\approx 10.5 $ GHz anti-crossings are due to parasitic coupling to circuit resonances. The qubit is prepared far off resonance with the cavity ($\nu_{\textrm{q}} = 10.116 $ GHz).}
\end{figure}

\subsubsection*{Methods - Cavity-qubit coupling}

\begin{figure}[ht]
\begin{center}
\includegraphics[scale = 3]{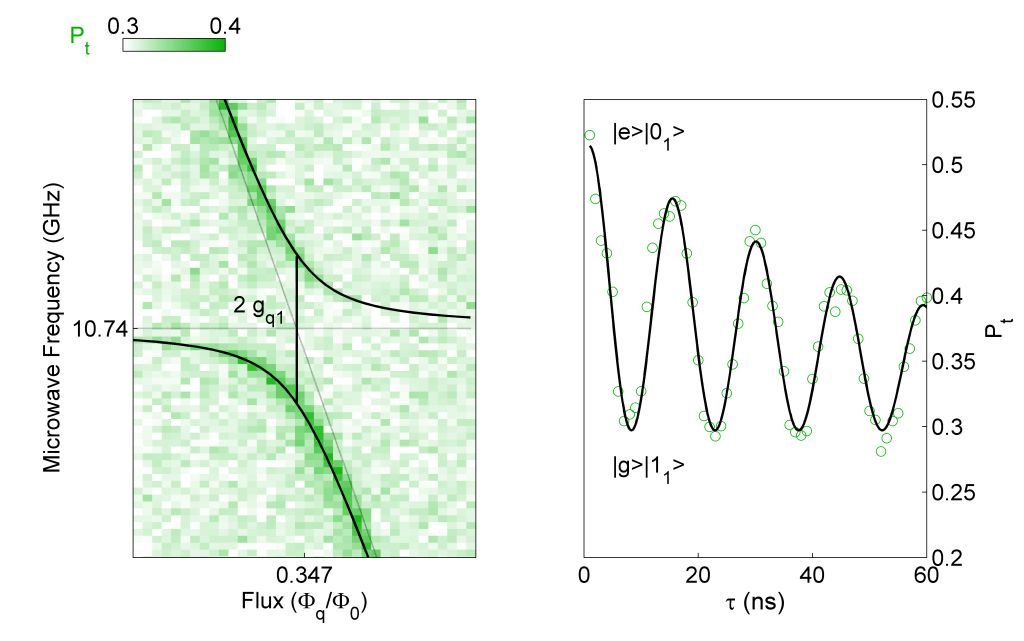}
\end{center}
\caption{3}{{\bf  Cavity-qubit coupling.} {\bf (Left)} Spectroscopy of the coupled system cavity-qubit. The tunneling probability of the qubit is plotted as a function of the qubit flux and driving frequency. It exhibits an avoided crossing when the qubit is tuned through the resonator frequency. A fit to the Jaynes-Cummings model is superposed over the data (black line), giving the value of the coupling rate $g_{\textrm{q}1} = $32 MHz. {\bf (Right)} Vacuum Rabi oscillations between the $\left| e \right>\left| 0_{1} \right>$ and the $\left| g \right>\left| 1_{1} \right>$ states when the qubit and the mode 1 of the cavity are in resonance during an interaction duration $\tau$.}
\end{figure}

Here we describe how the photon is loaded from the phase qubit into the cavity. The coupling between the qubit and the cavity is achieved by use of a coupling capacitor $C_{\textrm{g}} \approx $ 2.4 fF. This induces a coupling strength $g_{\textrm{q1}} = $ 32 MHz. Supplementary Figure 3 shows spectroscopy data of the cavity-qubit coupled system when the cavity is tuned to $\nu_1 = $ 10.74 GHz (point ``A'' on Supplementary Figure 5). On resonance, the dynamics of energy exchange between the qubit and the cavity is described, within the rotating wave approximation, by the Jaynes-Cummings Hamiltonian $[\hyperlink{33}{33}]$: $H_{\textrm{J-C}} = h g_{\textrm{q1}} (\hat{a}_{1}\hat{\sigma}_{+}+\hat{a}_{1}^{\dagger}\hat{\sigma}_{-})$, where $\hat{\sigma}_{+}$ and $\hat{\sigma}_{-}$ are the qubit raising and lowering operators. The qubit is prepared in its excited state far off resonance $\left( \left|\nu_{1}-\nu_{\textrm{q}}^{\textrm{off}}\right| >> 2 g_{\textrm{q1}}\right)$, then tuned in resonance with the cavity for an adjustable interaction duration $\tau$, and finally read out. During the cavity-qubit resonant interaction, the quantum of energy initially in the qubit oscillates between the two systems, \textit{i.e.}, between the states $\left| e \right>\left| 0_{1} \right>$ and $\left| g \right>\left| 1_{1} \right>$. In the absence of decoherence, this results in a cosine dependence of the qubit excited state occupancy, $P_{\left|e\right>}(\tau) = \frac{1+\cos (4\pi g_{\textrm{q1}} \tau)}{2}$ (Supplementary Figure 3). At half an oscillation period, the photon is completely transferred into the cavity, calibrating the $\pi$-pulse duration.

\subsubsection*{Methods - SQUID Model}
We consider a SQUID with a dc flux $\Phi_{\textrm{sq}}^{\textrm{dc}}$ and a current $i$ passing through it (Supplementary Figure 4). In the most general case, the SQUID can be modeled, for frequencies lower than its plasma frequency, as an inductance $L_{\textrm{sq}}(\Phi_{\textrm{sq}}^{\textrm{dc}},i)$, satisfying the voltage-current characteristic : $V = L_{\textrm{sq}}(\Phi_{\textrm{sq}}^{\textrm{dc}}, i) di/dt$ $[\hyperlink{34}{34}]$. We use an asymmetric SQUID intersected by two junctions J1 and J2 with critical currents $I_{\textrm{c}, 1}$ and $I_{\textrm{c}, 2}$ on one branch, and intersected by a junction J3 with a critical current $I_{\textrm{c}, 3}$ on the other branch. Junctions area are designed to give $I_{\textrm{c},2}\approx 2I_{\textrm{c},1}$, which means that J2 can be described as a pure geometric inductance $L_{2} = \varphi_0/I_{\textrm{c},2}$. J3 is designed to have a critical current such that $I_{\textrm{c},3} \approx 3 I_{\textrm{c},1}$. In this model we neglect the geometric inductance of the SQUID. We define (Supplementary Figure 4) the current $i_{1}$ (resp. $i_{3}$) through the junctions J1 (J3), and the reduced flux $\delta_{1}$ ($\delta_{3}$) across J1 (J3). The SQUID inductance is implicitly defined by the numerically solvable set of equations :
\begin{equation}
\left \{
\begin{array}{c}
i = i_{1} + i_{3} = I_{\textrm{c}, 1} \sin{\delta_{1}} + I_{\textrm{c}, 3} \sin{\delta_{3}} \\
\\
\delta_{3} - \delta_{1} - L_{2} \frac{I_{\textrm{c}, 1}}{\varphi_{0}} \sin{\delta_{1}} + 2\pi \frac{\Phi_{\textrm{sq}}^{\textrm{dc}}}{\Phi_{0}} = 0 \\
\\
\varphi_{0} L_{\textrm{sq}}^{-1}(\Phi_{\textrm{sq}}^{\textrm{dc}}, i) = \frac{d i}{d \delta_{3}}
\end{array}
\right..
\tag{S1}
\end{equation}
As we work at small intra-cavity power (for n=0, 1, 2, $\sqrt{\frac{h\nu_{n}}{L_{n}^{'}}} << I_{c, 1} + I_{c, 3}$), we neglect the dependence of the SQUID inductance with the current (we make the development of $L_{\textrm{sq}}$ at order 0 in $\delta_{3}^{\textrm{RF}}$ in Supplementary Equation 1).
\begin{figure}[ht]
\begin{center}
\includegraphics[scale = 0.5]{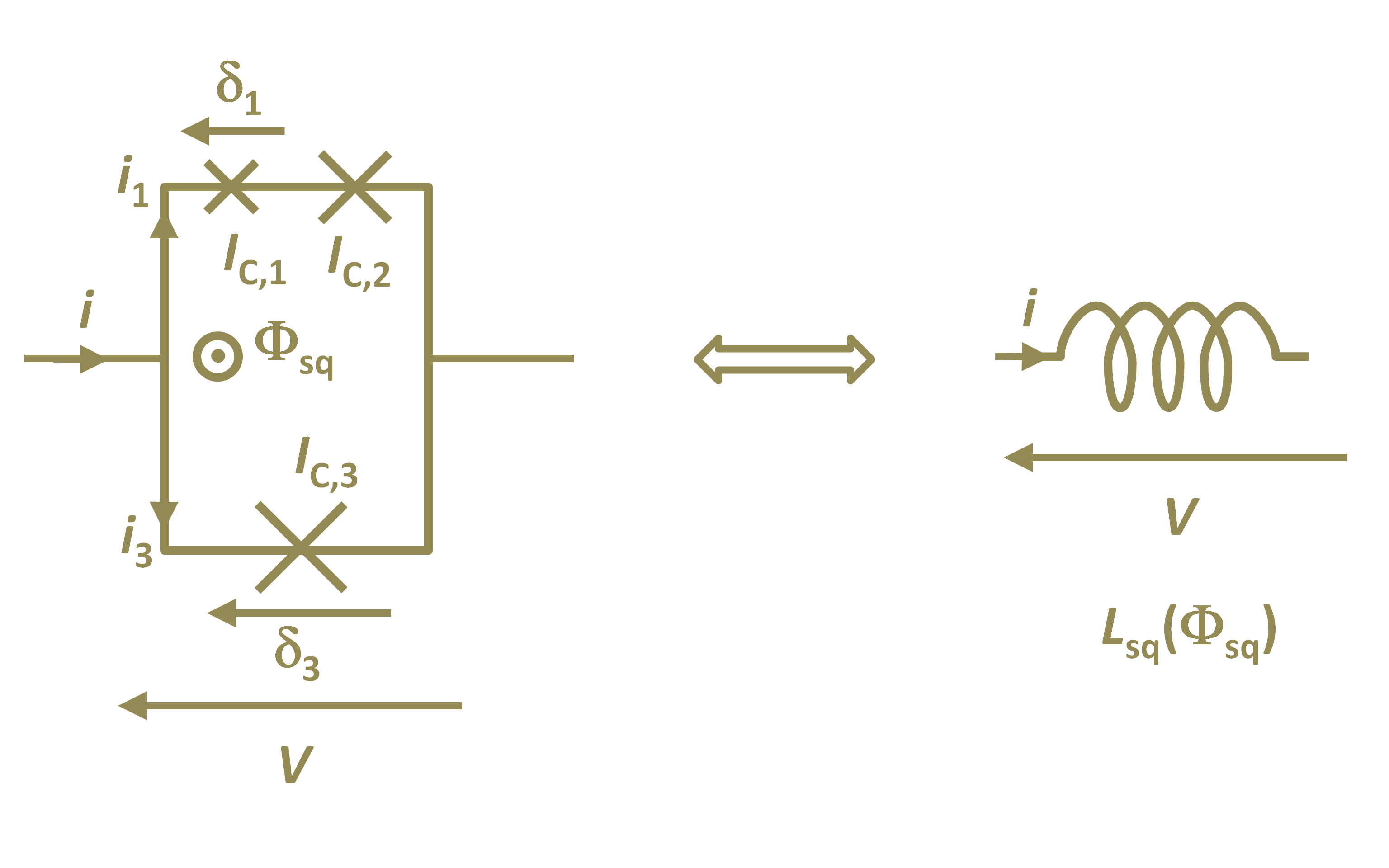}
\end{center}
\caption{4}{{\bf SQUID model.}}
\end{figure}

\subsubsection*{Methods - Fits to the resonance frequencies of the SPBS circuit}
To extract the SPBS circuit parameters, we use a more precise model than the one discussed in the main text. By writing the ``ABCD matrix" $[\hyperlink{35}{35}]$ of each component, we express the current $I$ and the voltage $V$ just before the coupling capacitance $C_{\textrm{c}}$ to the 50 $\Omega$ transmission line. A capacitance $C_{\parallel}$ in parallel with the SQUID, which includes the junction capacitances as well as a potential additional parasitic capacitance to ground, is taken into account. The characteristic impedances $Z_{1} = $ 50 $\Omega$ and $Z_{2} = $ 46 $\Omega$ are estimated by measuring the gap and the width of the CPW resonators with a scanning electron microscope. The total length of the CPW resonator corresponds to a bare resonance frequency of 4.14 GHz. Using the appropriate boundary condition, one gets
\begin{equation}
\Bigg[
  \begin{array}{c}
    V \\
    I
  \end{array}
\Bigg]
=
\Bigg[\begin{array}{c} C_{\textrm{c}} \end{array}\Bigg]
\Bigg[\begin{array}{c} \lambda/8(Z_1) \end{array}\Bigg]
\Bigg[\begin{array}{c} \lambda/8(Z_2) \end{array}\Bigg]
\Bigg[\begin{array}{c} L_{\textrm{sq}}^{\textrm{parallel}} \end{array}\Bigg]
\Bigg[\begin{array}{c} C_{\parallel} \end{array}\Bigg]
\Bigg[\begin{array}{c} V_{\textrm{sq}} \\ 0 \end{array}\Bigg]
\tag{S2}
\end{equation}
where $V_{\textrm{sq}}$ is the voltage across the SQUID. $I$ and $V$ are linked to the reflection coefficient $S_{11}$ following
\begin{equation}
\left[
  \begin{array}{c}
    1 \\
    S_{11}\\
  \end{array}
\right]
=
\left[
  \begin{array}{cc}
    1 & 1 \\
    1/50 & -1/50 \\
  \end{array}
\right]^{-1}
\left[
  \begin{array}{c}
    V/V^{+} \\
    I/V^{+} \\
  \end{array}
\right]
\tag{S3}
\end{equation}
where $V^{+}$ is the voltage of the incident wave on the capacitance $C_{\textrm{c}}$. For a given $\Phi_{\textrm{sq}}^{\textrm{dc}}$, the resonance frequencies are found by computing $S_{11}$ as a function of the frequency. Experimental results are compared with the theory by adjusting J1, J2, J3 critical currents and $C_{\parallel}$ (Supplementary Figure 5).
\begin{figure}[ht]
\begin{center}
\includegraphics[scale = 3]{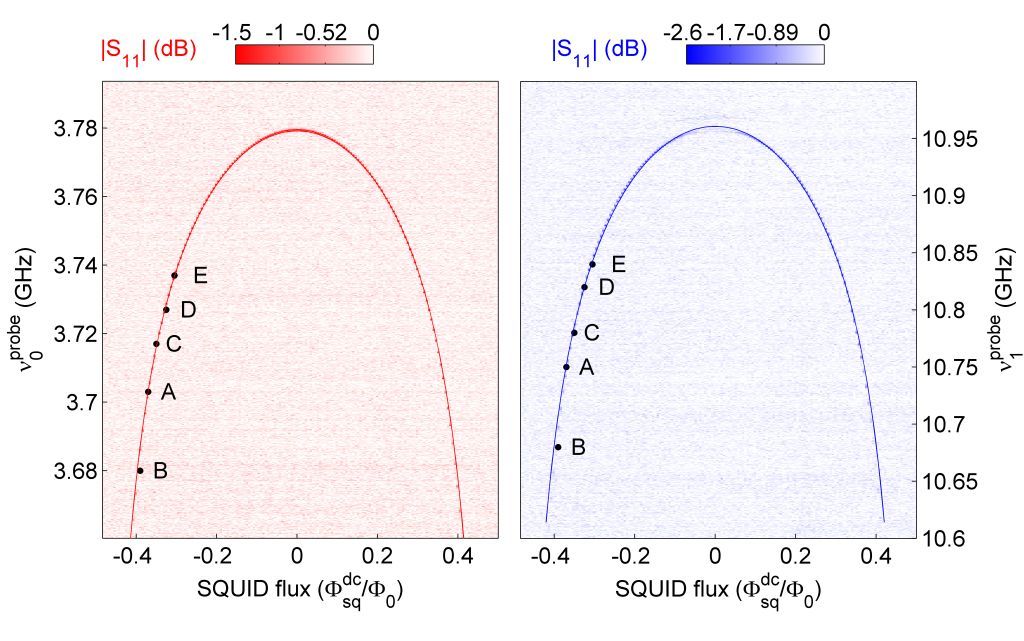}
\end{center}
\caption{5}{{\bf Fits to cavity spectroscopies.}}
\end{figure}

\subsubsection*{Discussion - Comments on the flux expansion of the SQUID inductance}
For the sake of simplicity, the inductance of the SQUID has been expanded only to first order in the flux ($\Phi_{\textrm{sq}}^{\mu \textrm{w}}(t)$). We find that this approximation adequately describes the experimental results. Here we explain the effects of the higher order terms of the SQUID inductance expansion. We limit this discussion to the third order expansion (this can be easily generalized to higher order). We write
\begin{equation}
\frac{L_{\textrm{sq}}(\Phi_{\textrm{sq}}^{\textrm{dc}} + \Phi_{\textrm{sq}}^{\mu\textrm{w}}(t))}{L_{\textrm{sq}}(\Phi_{\textrm{sq}}^{\textrm{dc}})} = 1 + \delta L_{\textrm{sq}}^{(1)}
\cos(2\pi \nu_{\textrm{p}} t + \varphi_{\textrm{p}}) +   \delta L_{\textrm{sq}}^{(2)}
\cos^{2}(2\pi \nu_{\textrm{p}} t + \varphi_{\textrm{p}}) + \delta L_{\textrm{sq}}^{(3)}
\cos^{3}(2\pi \nu_{\textrm{p}} t + \varphi_{\textrm{p}})
\tag{S4}
\end{equation}
with
\begin{equation}
\delta L_{\textrm{sq}}^{(i)} = \frac{\partial^{i}(L_{\textrm{sq}}/L_{\textrm{sq}}(\Phi_{\textrm{sq}}^{\textrm{dc}}))}{\partial (\Phi/\Phi_{0})^i} \Bigg|_{\Phi_{\textrm{sq}}^{\textrm{dc}}}\frac{1}
{i!} \left( \frac{\delta \Phi_{\textrm{sq}}^{\mu \textrm{w}}}{\Phi_{0}} \right)^{i}.
\tag{S5}
\end{equation}
Gathering terms oscillating at the same frequency, we get
\begin{equation}
\frac{L_{\textrm{sq}}(\Phi_{\textrm{sq}}^{\textrm{dc}} + \Phi_{\textrm{sq}}^{\mu\textrm{w}}(t))}{L_{\textrm{sq}}(\Phi_{\textrm{sq}}^{\textrm{dc}})}  = C_{0} + C_{1} \cos(2\pi \nu_{\textrm{p}} t + \varphi_{\textrm{p}}) + C_{2} \cos(2(2\pi \nu_{\textrm{p}} t + \varphi_{\textrm{p}})) + C_{3} \cos(3(2\pi \nu_{\textrm{p}} t + \varphi_{\textrm{p}}))
\tag{S6}
\end{equation}
with
\begin{eqnarray}\nonumber
C_{0} &=& 1 + \frac{1}{2}\delta L_{\textrm{sq}}^{(2)}\nonumber \\
C_{1} &=& \delta L_{\textrm{sq}}^{(1)} + \frac{3}{4} \delta L_{\textrm{sq}}^{(3)}\nonumber \\
C_{2} &=& \frac{1}{2}\delta L_{\textrm{sq}}^{(2)}\nonumber \\
C_{3} &=& \frac{1}{4} \delta L_{\textrm{sq}}^{(3)}\nonumber.
\end{eqnarray}
We now compare this expansion with the simple first-order approximation used in the report. To illustrate our discussion, we have plotted in Supplementary Figure 6 the $\delta L_{\textrm{sq}}^{(i)}$, $ i\in\{1, 2, 3\}$, terms as a function of the flux bias, for a reference microwave flux modulation $\delta \Phi_{\textrm{sq}}^{\mu\textrm{w}} = $ 0.02 $\Phi_{0}$. In the following, we calculate the effect of the higher-order terms at flux bias equal to -0.37 $ \Phi_{0}$ (point ``A'' on Supplementary Figure 5). The dc rectification of the SQUID inductance is about 0.2 $\%$. This corresponds to a relative frequency shift $\sim$ 0.02 $\%$, which is in rough agreement with the -5 MHz frequency shift measured experimentally for the n = 1 resonance. The rectification to the term oscillating at $\nu_{\textrm{p}}$ is about 1$\%$, which is reasonable to neglect at our level of precision on the experimental parameters (particularly on the estimation of the pump amplitude at the circuit input). The term oscillating at 2$\nu_{\textrm{p}}$ corresponds to a ``two pump photons" parametric interaction that could, in a $\lambda /4$ cavity with the usual $\nu_{\textrm{n}} = (2n+1)\nu_{0}$ dispersion, couple the mode n with the mode n+2. Our main goal has been to isolate the 0$\leftrightarrow$1 conversion. Experimentally, we measure $(\nu_{2} - \nu_{0}) - 2(\nu_{1}-\nu_{0})= $ 240 MHz, which is much higher than both the bandwidths of the resonances and the coupling rates involved. We also see no trace of the 1$\leftrightarrow$3 conversion while pumping 1$\leftrightarrow$0. Note that the splitting signature of these two pump photons processes is different, the ``quasi-resonant pump" conditions being expressed as $2\nu_{\textrm{p}} = \nu_{\textrm{n+2}} - \nu_{\textrm{n}}^{\textrm{probe}}$ or $2\nu_{\textrm{p}} = \nu_{\textrm{n}} + \nu_{\textrm{n+2}}^{\textrm{probe}}$. In addition to that, it requires a stronger pump amplitude to achieve the same coupling rate as the single pump-photon transition (for example, $g_{20} \sim$ 0.08 $g_{10}$). Processes involving three pump photons can be treated the same way, but the effect is even smaller.
\begin{figure}[ht]
\begin{center}
\includegraphics[scale = 0.5]{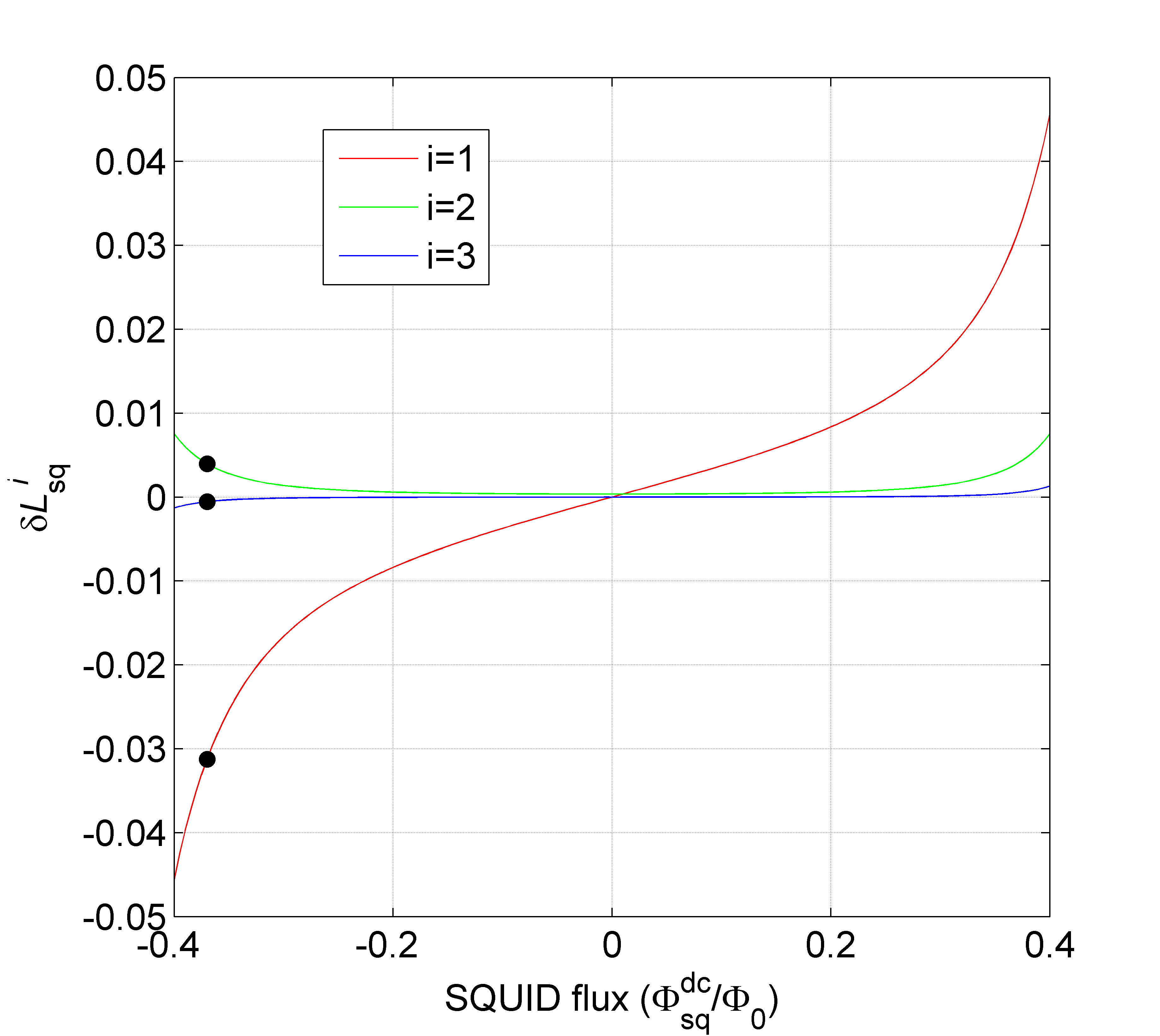}
\end{center}
\caption{6}{{\bf Higher order terms of the SQUID inductance expansion with flux.} Plot of the $\delta L_{\textrm{sq}}^{(i)}$ factors of Eq. (5), for $i \in \left\{1, 2, 3\right\}$,  as a function of the normalized flux $\Phi_{\textrm{sq}}^{\textrm{dc}}/ \Phi_{0}$. Dark points represent the flux point ``A'' on Supplementary Figure 5 (-0.37 $\Phi_{0}$).}
\end{figure}

\clearpage
{\Large {\bf References}}

\renewcommand{\theenumi}{\arabic{enumi}}
\begin{enumerate}
\setcounter{enumi}{30}
\item  \hypertarget{31}{ }Neeley, M. {\it et al.} Transformed dissipation in superconducting quantum circuit. {\it Phys. Rev. B} {\bf 77}, 180508 (2008).
\item  \hypertarget{32}{ }Simmonds, R.W. {\it et al.} Decoherence in Josephson phase qubits from junction resonators. {\it Phys. Rev. Lett.} {\bf 93}, 77003 (2004).
\item  \hypertarget{33}{ }Jaynes, E. \& Cummings, F. Comparison of quantum and semiclassical radiation theories with application to the beam maser. {\it Proc IEEE} {\bf 51}, 89 (1963).
\item  \hypertarget{34}{ }Clarke, J. \& Braginski, A.I. {\it The SQUID Handbook} (Wiley-Vch, USA, 2006).
\item  \hypertarget{35}{ }Pozar, D. {\it Microwave Engineering} (Wiley-India, 2009).
\end{enumerate}

\end{document}